\documentclass[preprint,5p,times,twocolumn]{elsarticle}

\usepackage{amssymb}
\usepackage{units}

\usepackage{lineno}
\journal{Nuclear Instruments and Methods, Research Section A}

\begin{document}

\begin{frontmatter}



\title{A Method to Determine the Maximum Radius of Defocused Protons after Self-Modulation in AWAKE}

\author[CERN]{M. Turner}
\author[CERN]{E. Gschwendtner}
\author[CERN,MPP]{P. Muggli}
\address[CERN]{CERN, Geneva, Switzerland}
\address[MPP]{Max-Planck Institute for Physics, Munich, Germany}

\begin{abstract}
The AWAKE experiment at CERN aims to drive GV/m plasma wakefields with a self-modulated proton drive bunch, and to use them for electron acceleration. During the self-modulation process, protons are defocused by the transverse plasma wakefields and form a halo around the focused bunch core. The two-screen setup integrated in AWAKE measures the transverse, time-integrated proton bunch distribution downstream the \unit[10]{m} long plasma to detect defocused protons. By measuring the maximum radius of the defocused protons we attempt calculate properties of the self-modulation. In this article, we develop a routine to identify the maximum radius of the defocused protons, based on a standard contour method. We compare the maximum radius obtained from the contour to the logarithmic lineouts of the image to show that the determined radius identifies the edge of the distribution.
\end{abstract}

\begin{keyword}
AWAKE \sep  Seeded Self-Modulation \sep Plasma Wakefield Acceleration \sep Two-Screen Setup



\end{keyword}

\end{frontmatter}


\section{Introduction}
\label{sec:intro}
\begin{figure}[htb!]
\centering
		\includegraphics[width =1\columnwidth]{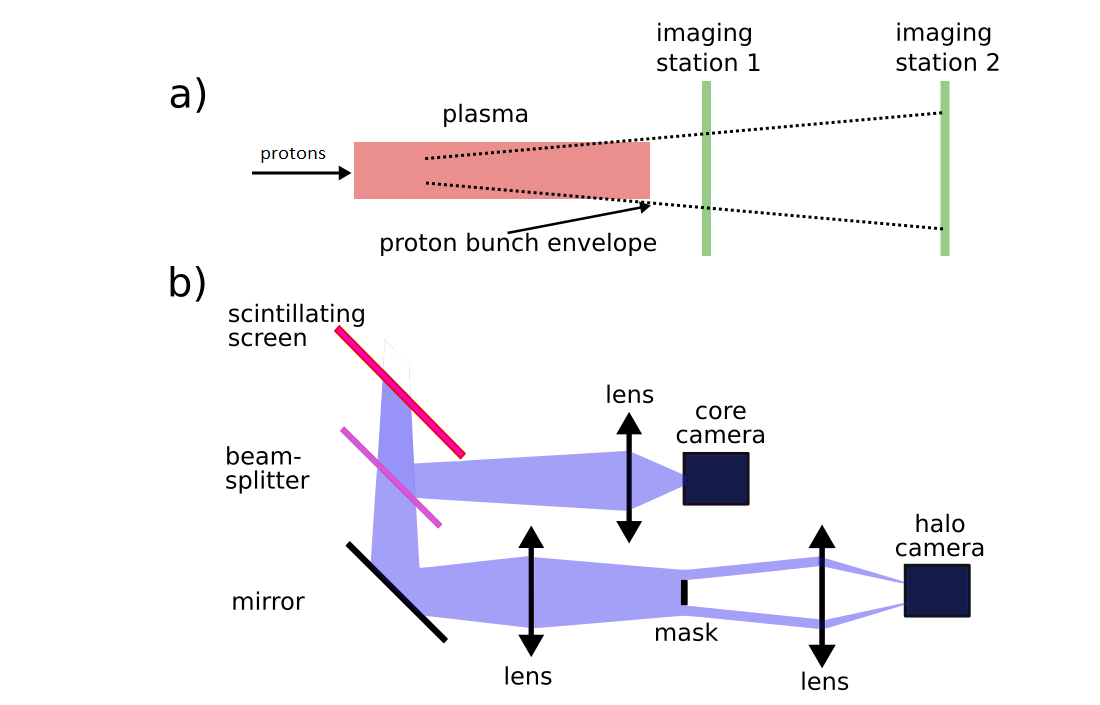}
		\caption{a) Schematic location of the imaging stations with respect to the plasma. The proton bunch moves to the right. b) Schematic drawing of the optics of the imaging stations. Drawings are not to scale.}
\label{fig:schematic}
\end{figure}

The Advanced Proton Driven Plasma Wakefield Experiment (AWAKE) \cite{AWAKE} at CERN aims to use plasma wakefields to accelerate \unit[10-20]{MeV} electrons to GeV level in \unit[10]{m} of plasma. The wakefields are created by a self-modulated proton drive bunch. To show that the proton bunch self-modulates, the experiment uses the two-screen measurement setup \cite{TWOSCREEN1,TWOSCREEN2}, which detects protons that are defocused by the transverse plasma wakefields. By identifying the maximum radius of the defocused proton distribution we can learn about the properties and development of the self-modulation. The two-screen setup consists of two imaging stations that measure the transverse, time integrated proton bunch distribution downstream the end of the plasma (see Figure \ref{fig:schematic}a).

Figure \ref{fig:schematic}b shows a schematic drawing of the imaging stations. Each imaging station uses a scintillating Chromox (Al$_2$O$_3$:CrO$_2$) screen and two cameras, that are synchronized to the bunch arrival time. When the bunch traverses the scintillating screen, it deposits energy in the material and the material emits photons in the visible range. The number of emitted photons is proportional to the energy deposit in the material\footnote{The energy-loss of a proton in the \unit[1]{mm} thick Chromox screen  depends on its energy. In our case their incoming energy is \unit[400]{GeV} and we expect the wakefields to change the proton-energy by less than \unit[$\pm 10$]{GeV}. The proton energy-loss in the screen material is approximately \unit[1]{MeV}, which is much smaller than \unit[400]{GeV}. All proton energies are within a narrow range. It is therefore reasonable to assume that their energy loss is the same.}, which is proportional to the number of protons \cite{MYTHESIS}. We then image the emitted light onto cameras.

From simulations we expect that for each strongly defocused proton, we get approximatmely \unit[$10^4-10^5$]{protons} that were not defocused and stay in the bunch core \cite{TWOSCREEN1,MYTHESIS}. To detect both the maximum defocused protons and the ones in the core, we use two cameras (see Figure \ref{fig:schematic}b). 

We split the light emitted by the scintillating screen with a beam-splitter. One part is directly imaged by a lens onto the first camera, which we call the core camera (Basler ac1300-30gm). The second part is imaged onto a \unit[1]{mm} thick glass window (mask) with round opaque coatings that have an optical density larger than 5 and block the light emitted by the bunch core. The light passing around the coatings is then imaged by another lens onto the second camera, which we call the halo camera (Basler ac1600-60gm).

\begin{figure}[htb!]
\centering
		\includegraphics[width =0.8\columnwidth]{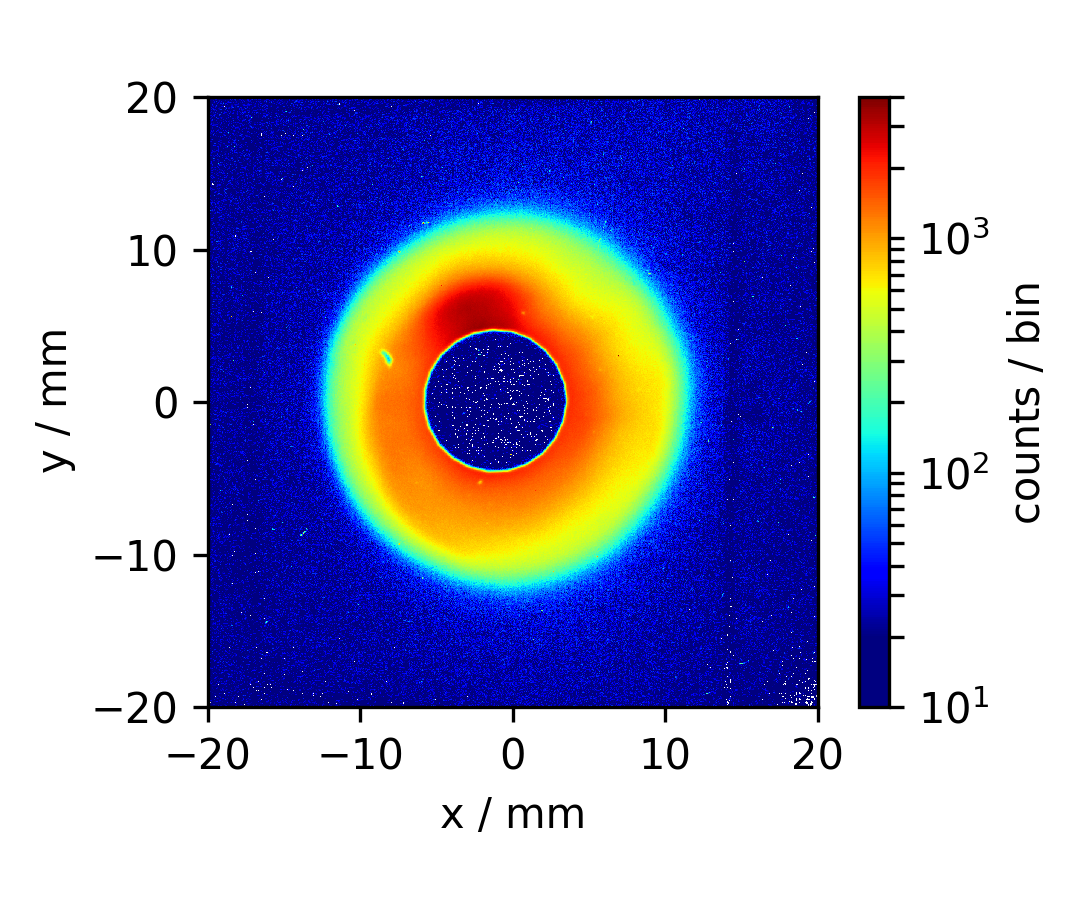}
		\caption{Typical image of a defocused proton distribution measurement by the halo camera of one of the imaging stations. Note the logarithmic scale. The light of the bunch core is blocked by a mask.}
\label{fig:7e14image}
\end{figure}

In this article we describe a routine that enables us to determine the maximum radius of the proton distribution on the halo camera images. This maximum radius is the basis for all further calculations. A typical measurement of the defocused proton distribution taken by one of the halo cameras is shown in Figure \ref{fig:7e14image}. Note that the core of the bunch is blocked by the mask, and that the distribution has a distinct outer edge.

\section{Method}
\label{sec:method}

In order to determine the maximum defocused proton bunch radius, we put together a procedure that (semi-) automatically determines the largest radius at which light and thus protons are detected on the halo camera images. This procedure assumes -as observed in simulations \cite{TWOSCREENCONCEPT}- that the defocused proton halo has a distinct maximum radius. The method uses a standard contour subroutine for 2D images\footnote{Python: contour-function from the matplotlib package}.
We perform the following steps on the images:

1) We apply a median filter \cite{MEDIANFILTER} to remove pixels with high counts due to secondary particles impacting on the camera chip and to pixel upset. We verify that the median filter does not change the transverse profile, especially close to the edge of the distributions.

2) To obtain the centroid of the bunch distribution of each halo camera image, we fit the horizontal and vertical projections of the corresponding core camera image with a Gaussian distribution and obtain the center of the peak. Since we previously obtained the scaling between the core and halo cameras from measurements without plasma or mask, we now use this information to determine the proton bunch core centroid on each halo camera image, with mask.

3) Calculation of contours of the counts distribution of the halo cameras.

4) Selection of the longest closed outer contour above background level around the signal as the edge of the halo. When the selected level of the contour is chosen too low, the contour is not closed around the center of the bunch and is very short. When the contour level is set too high, the contour does not represent the maximum defocused radius and is shorter. 
The height of this level is selected and tested on one image of each imaging station and of each series, then tested and cross-checked with several images of the measurement series (i.e. measurement performed with the same camera settings) and then applied to each measurement of the series. Afterwards, every contour of every image is verified by eye and the maximum radius that was selected is compared to the images lineouts and projections.

5) Calculation of the maximum proton bunch radius from the distance of each point of the contour to the bunch center position. The maximum radius for each image is calculated as the mean of the radii of each contour point. We use the standard deviation of the measured radii to characterize the variation in determining the maximum defocused proton distribution radius.

Note that the uncertainty of the maximum is determined by the resolution of the setup as well as the asymmetry of the measured signal\footnote{Because of cylindrical symmetry in the process, the signal is expected to be symmetric. Asymmetries can be caused by the misalignment between the proton bunch and the plasma and can increase the uncertainty on the maximum radius of the distribution.}.

\section{Results}
\begin{figure}[htb!]
\centering
		\includegraphics[width =\columnwidth]{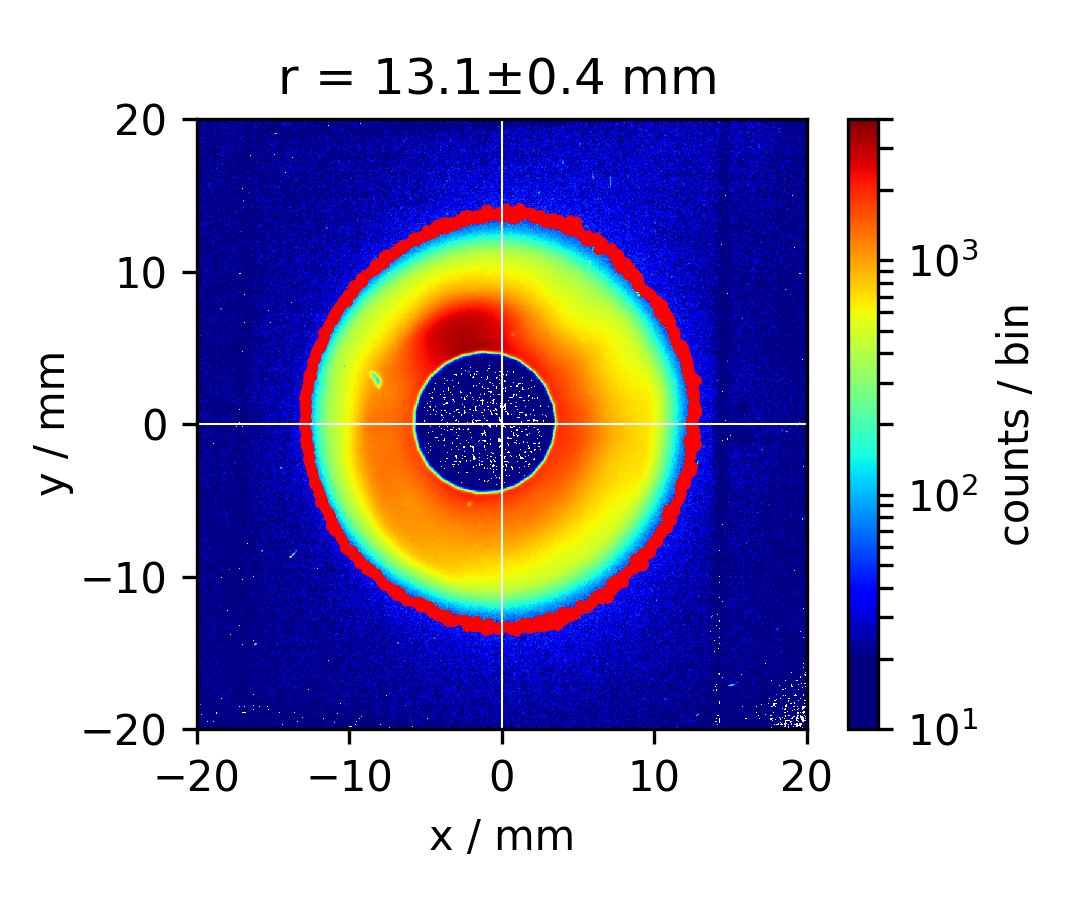}
		\caption{Same measurement image as in Figure \ref{fig:7e14image}, where the calculated contour is shown in red. The center of the proton bunch core is marked by a white cross. Note the logarithmic scale.}
\label{fig:7e14contour}
\end{figure}

We calculate the maximum proton bunch radius of Figure \ref{fig:7e14image} based on the procedure described in Section \ref{sec:method}. Figure \ref{fig:7e14contour} shows the contour that was determined as being the edge of the distribution (red)  at the same time as the from the core camera image determined proton bunch center (white cross). In this case, the average radius is \unit[13.1]{mm} and the standard deviation of the radii along the contour is \unit[0.4]{mm}.

\label{sec:results}

\begin{figure}[htb!]
\centering
		\includegraphics[width =\columnwidth]{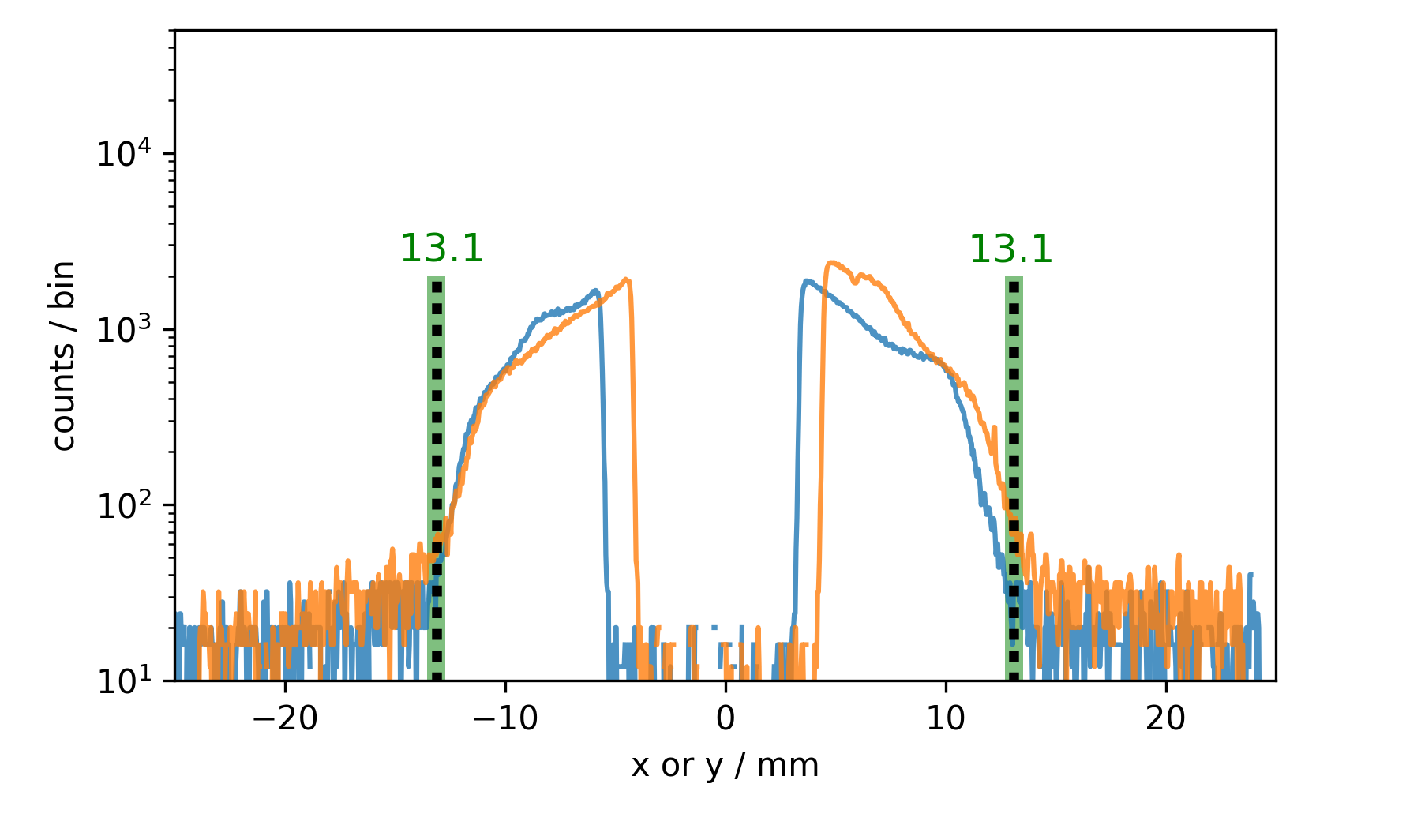}
		\caption{Horizontal (blue) and vertical (orange) lineouts of Figure \ref{fig:7e14contour}. The maximum radius determined with the contour procedure is shown as the black vertical dashed lines and the standard deviation as the green thick line.}
\label{fig:7e14comp}
\end{figure}

To verify that the maximum radius of the proton distribution makes sense, we compare on a logarithmic scale the radius we obtain for the maximum defocused edge to the horizontal and vertical projection of the measurement. Figure \ref{fig:7e14comp} shows the maximum proton bunch radius (vertical bars where the green area marks the uncertainties) as well as the horizontal and vertical projections of the measurement shown in Figure \ref{fig:7e14contour}. 

The lineouts show that the recorded distributions have a clear edge and that the maximum defocused proton bunch radius determined by the contour method indeed identifies the largest radius of the distribution above background level.  

\section{Conclusions}

For this diagnostic, the AWAKE experiment at CERN is interested in the protons that are defocused by the transverse plasma wakefields, and in their maximum radial position at the two imaging stations. Thus, in this article we present a method to determine the maximum radius of a measured proton distribution with a distinct maximum radius. The method uses a standard Python contour routine and selects the longest closed contour above background level, closed around the core of the distribution. From this contour, we can calculate the mean radius from the bunch center and its standard deviation. The bunch center is determined by Gaussian fits from the images of the cameras recording the bunch core, separately for each measurement. By comparing the maximum radius to the lineouts and profiles of the measurements, we show that the calculated radius identifies the edge of the proton distribution.

\section{Acknowledgements}
The authors would like to thank the Beam Instrumentation Group at CERN, which greatly helped with the design and installation of the imaging stations in the AWAKE facility and we acknowledge the contributions from the AWAKE experimental team.

\appendix

\end{document}